\documentclass[11pt]{article}

\usepackage{amsfonts}
\usepackage{appendix}
\usepackage{verbatim}

\usepackage{amsmath}
\usepackage{amssymb}
\usepackage{latexsym}
\usepackage{graphicx}
\usepackage[english]{babel}
\usepackage[font={small}]{caption}
\usepackage{mathtools}
\usepackage{geometry}                
\usepackage{epstopdf}
\usepackage{tikz-cd}
\usetikzlibrary{cd}
\usepackage{amsfonts}
\usepackage{latexsym}
\usepackage{graphicx}
\usepackage[english]{babel}
\usepackage{tikz}
\usepackage{soul}
\allowdisplaybreaks

\usepackage{mciteplus}

\usepackage[colorlinks=true,      linkcolor=blue,      urlcolor=blue,      filecolor=blue,      citecolor=blue,
      pdfstartview=FitH,     pdfpagemode=UseNone,      bookmarksopen=true]{hyperref}

\usepackage{subcaption}
\begin{document}
\input epsf

\def\p{\partial}
\def\h{{1\over 2}}
\def\be{\begin{equation}}
\def\bea{\begin{eqnarray}}
\def\ee{\end{equation}}
\def\eea{\end{eqnarray}}
\def\d{\partial}
\def\la{\lambda}
\def\eps{\epsilon}
\def\b{\bigskip}
\def\m{\medskip}

\renewcommand{\a}{\alpha}	
\renewcommand{\b}{\beta}
\newcommand{\g}{\gamma}		
\newcommand{\G}{\Gamma}
\renewcommand{\d}{\delta}
\newcommand{\D}{\Delta}
\renewcommand{\c}{\chi}			
\newcommand{\C}{\Chi}
\renewcommand{\P}{\Psi}
\newcommand{\s}{\sigma}		
\renewcommand{\S}{\Sigma}
\renewcommand{\t}{\tau}		
\newcommand{\e}{\epsilon}
\newcommand{\n}{\nu}
\renewcommand{\r}{\rho}
\renewcommand{\l}{\lambda}

\newcommand{\newsection}[1]{\section{#1} \setcounter{equation}{0}}

\def\q{\quad}

\def\h{{1\over 2}}
\def\t{\tilde}
\def\r{\rightarrow}
\def\nn{\nonumber\\}

\let\p=\partial

\newcommand\blfootnote[1]{%
  \begingroup
  \renewcommand\thefootnote{}\footnote{#1}%
  \addtocounter{footnote}{-1}%
  \endgroup
}

\begin{flushright}
\end{flushright}
\vspace{20mm}
\begin{center}
{\LARGE The Dual of a Tidal Force in the D1D5 CFT}
\\
\vspace{18mm}
\textbf{Bin} \textbf{Guo}{\footnote{guo.1281@buckeyemail.osu.edu}}~\textbf{and} ~ \textbf{Shaun}~  \textbf{Hampton}{\footnote{shaun.hampton@ipht.fr}}
\\
\vspace{10mm}
${}^1$Department of Physics,\\ The Ohio State University,\\ Columbus,
OH 43210, USA\\ \vspace{8mm}

${}^2$Institut de Physique Th\'eorique,\\
	Universit\'e Paris-Saclay,
	CNRS, CEA, \\ 	Orme des Merisiers,\\ Gif-sur-Yvette, 91191 CEDEX, France  \\

\vspace{8mm}
\end{center}

\vspace{4mm}

\thispagestyle{empty}
\begin{abstract}
It was demonstrated that a string probe falling radially within a superstratum geometry would experience tidal forces. These tidal forces were shown to excite the string by converting its kinetic energy into stringy excitations. Using the AdS/CFT correspondence we seek to understand this behavior from the perspective of the dual D1D5 CFT. To study this process we turn on an interaction of the theory which is described by a deformation operator. We start with an initial state which is dual to a graviton probe moving within the superstratum geometry. We then use two deformation operators to compute transition amplitudes between this state and a final state that corresponds to stringy excitations. We show that this amplitude grows as $t^2$ with $t$ being the amount of time for which the deformation operators are turned on. We argue that this process in the CFT is suggestive of the tidal effects experienced by the probe propagating within the dual superstratum geometry.

\vspace{3mm}

\end{abstract}
\newpage

\setcounter{page}{1}

\numberwithin{equation}{section} 

\tableofcontents

\newpage

\section{Introduction}

The microstate geometry program \cite{fuzzballs_i,fuzzballs_ii,fuzzballs_iii,fuzzballs_iv,fuzzballs_v} has enjoyed much progress in recent years. In this paradigm the black hole is replaced by a smooth, horizonless geometry of similar mass and charge constructed from objects in string theory. It is difficult to distinguish these geometries from classical black holes from far away. As an observer approaches the would be horizon, the microstructure is resolved and the geometry caps off smoothly in all known examples. Many solutions of various charges have been constructed. The two charge microstates are built from BPS configurations of D1 and D5 branes wrapping compact directions and were first constructed in \cite{Lunin:2001fv}. 
More recently, fully back reacted three charge microstates have been constructed by placing a left moving momentum wave along the common direction of the wrapped D1 and D5 branes. These geometries are known as superstrata \cite{Bena:2015bea,bmtw,bgmrstw2,bgmrstw,Ceplak:2018pws,hw}.  A very nice feature is that they have an asymptotic region which is $AdS_3\times S^3\times T^4$ and using the AdS/CFT correspondence \cite{maldacena,gkp,witten} they admit a precise map between them and states in the dual D1D5 CFT \cite{vev}. In this paper we will focus on the CFT dual of a particularly simple class of geometries known as the $(1,0,n)$ superstrata.

One way to elucidate properties of the geometry under consideration is to send in a probe and analyze how it is affected. One such effect which can be studied is the tidal force. Tidal forces arise due to curvature effects which, for an extended object, act differently at different positions along the object. For example, consider a classical Schwarzschild black hole. The scalar curvature, $R$, is zero. However, some components of the curvature tensor are nonzero but in the end must cancel when contracted in the appropriate way. This is essentially because you have a stretching of space along the radial direction and a compression of space along the directions which are transverse. These curvature affects act on an infalling observer which is extended in space. 
In \cite{tww,bmww,bhw}, it was shown that, for superstrata geometries, tidal forces become significant at a macrosopic distance from the cap. These deviations from classical black hole geometries at macroscopic distances from the cap may help provide observational signatures of these objects. See \cite{dm} for a review of observational effects. In \cite{mw}, an infalling graviton probe was analyzed within an explicit realization of a superstratum geometry. 
The radial kinetic energy was converted into stringy excitations along the common circle of the wrapped D1D5 branes and a direction along $S^3$.
As the string propagated towards the center of the geometry and back up the other side of the throat it wasn't able to make it back to the AdS boundary. This is due to the loss of kinetic energy into stringy excitations. As the string yo-yo'ed back and forth within the throat each successive return height was less than the previous one. The string became tidally trapped.  
In \cite{chl} the authors showed that the string is also tidally excited along directions of $T^4$. For other works on tidal forces in the context of AdS/CFT see \cite{stretching,craps,Engelsoy:2021fbk}.

In this paper we are interested in understanding these tidal effects from the perspective of the dual CFT. The D1D5 system has proved extremely useful in studying microscopic features of black holes such as the counting of black hole microstates \cite{sv}, reproducing greybody factors in Hawking radiation \cite{cm}. Much about microstate geometries can be learned from the the D1D5 CFT at the orbifold point where the theory is free \cite{lm1,lm2,dmw,Larsen:1999uk,orbifold2,Arutyunov:1997gt,Burrington:2012yq,taylor,Dei:2019iym}. However, if one wants to understand dynamical processes in supergravity, such as tidal effects and the scrambling of probes one must turn on an interaction in the CFT \cite{Avery:2010er,Burrington:2014yia,Carson:2016uwf,peet}. This is done by deforming away from the free point by a marginal deformation which includes a twist operator. This operator joins and splits `effective component strings' or `strands' that wrap the common circle of the D1D5 bound state. 

The CFT dual of a free falling graviton in AdS was studied in the context of the D1D5 CFT \cite{hm,dissertation,Guo:2021ybz}. This involves applying two marginal deformations to a left and a right moving mode in the initial state which corresponds to a \textit{single} graviton propagating in the gravity dual. Transition amplitudes were then computed which describe the likelihood of the left and right moving mode to split apart into more modes of lower energy in the final state. The nature of this splitting mechanism as a function of time helps describe what is happening in the gravity dual. We considered the simplest splitting process in the scenario: one left and one right moving mode splitting into three left and three right moving modes in the final state. We call this the $1\to 3$ process. At early times, the amplitude is linear in time.
This corresponds to an infalling graviton becoming redshifted in AdS.
At late times, the amplitude is periodic. The oscillatory nature indicated that a graviton which started from the boundary, propagated to the center of $AdS$ to the boundary on the other side and then back again to its original position. This is expected since a graviton moving in $AdS$ is traveling along a geodesic, it isn't expected to be thermalized.

In \cite{hm,dissertation} we also computed a $2\to 4$ process: two left and two right moving modes splitting into four left and four right moving modes in the final state.
The amplitude grows like $t^2$ instead of being periodic as in the $1\to 3$ process. 
This growing behavior is a preliminary signal of thermalization. 
This would correspond to two gravitons colliding in the gravity dual. 
On the other hand, in \cite{mw} the tidal force on an infalling graviton is studied in the $(1,0,n)$ superstratum geometry. The infalling graviton would become tidally excited into string states. In the CFT dual, the state corresponding to the $(1,0,n)$ superstratum is in the Ramond sector. It has left moving modes with nonzero energy and right moving modes with zero energy. If we consider the splitting of one left and right mover in this CFT background, we expect that the left and right mover will collide with extra left and right movers in the background. This could result in a growing term like $t^2$ as in the  $2\to 4$ process and may explain the tidal force in \cite{mw}.

\begin{figure}[!ht]
\centering
        \includegraphics[width=9cm]{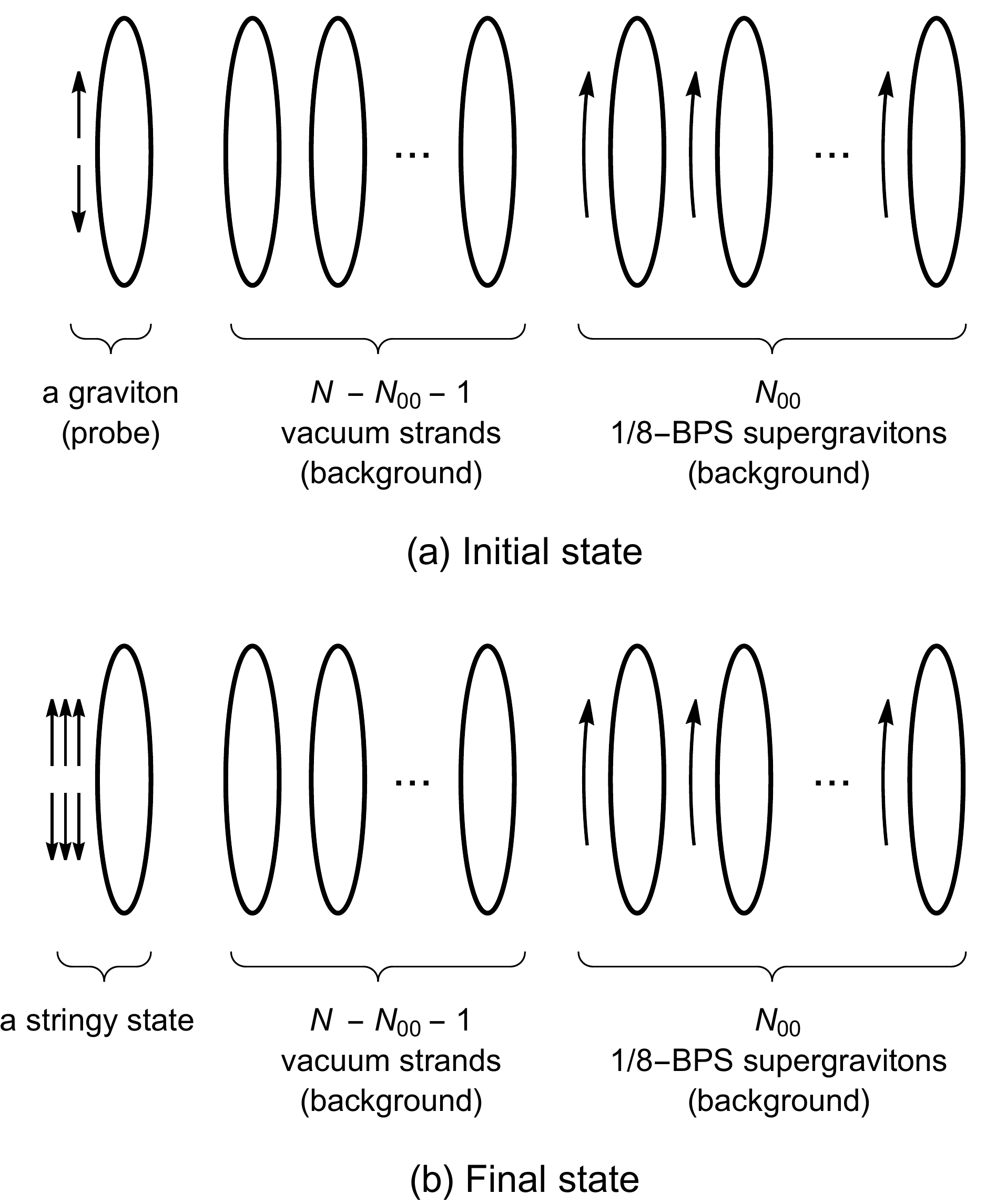}
\caption{The initial and final state of the splitting process within the background superstratum state. In the gravity dual, this corresponds to a probe graviton being tidally excited to a string state by the superstratum geometry.}
\label{fig_states}
\end{figure}

In this paper, within the CFT, we apply two deformations to one left and one right mover in the initial state, built not on the vacuum, but on the CFT dual of a superstratum geometry which we call a \textit{superstratum} state. We consider the splitting of this initial probe into three left and three right movers in the final state with the superstratum state remaining unaltered. This is done to remove any effects coming from the backreaction of the geometry. The initial and final states are depicted in fig.\,\ref{fig_states}. In this case, we show that though the splitting is $1\to3$ we still find a quadratic growth in the amplitude in time $t$. The same behavior which marked the $2\to4$ process computed in the vacuum state \cite{hm}. The final state containing three modes resembles a single particle stringy state in the superstratum geometry. We therefore see the effect of the superstratum state on the type of splitting behavior that is produced. This presents evidence, within the CFT, for the tidal excitations of an infalling probe created by the superstratum geometry in the gravity dual.

We outline the paper as follows. In section \ref{sec 2} we review the D1D5 CFT and the deformation operator. In section \ref{sec 3} we describe the splitting process and construct the initial and final states shown in fig.\,\ref{fig_states}. In section \ref{sec 4} we outline the computation of the splitting amplitude. We find that there exists a $t^2$ secular term corresponding to tidal forces. In section \ref{sec 5} we numerically compute the amplitude for various initial energies and superstrata momenta. By extrapolating the numerical results, we find an analytical expression for the amplitude in the limit where the initial energy (initial radial momentum of the probe graviton) is large. In section \ref{sec 6} we study the amplitude in the large $N$ limit. Finally, in section \ref{sec 7} we discuss our results and outlook.

\section{The D1D5 CFT}\label{sec 2}

In this section we summarize the main features of the D1D5 CFT which we use in this computation. We also describe the deformation operator which perturbs us away from the free theory. We start with Type IIB string theory compactified on
\bea
M_{9,1}\to M_{4,1}\times S^1\times T^4
\eea
Now wrap $N_1$ D1 branes around $S^1$, and $N_5$ D5 branes around $S^1\times T^4$. The D1 and D5 branes coincide along $S^1$. We consider the $S^1$ to be much larger than the $T^4$ so that at low energies the bound state of this system describes a 2D CFT. 
It has been conjectured that one can move in the moduli space of string couplings to a point where the theory is free. This is called the `orbifold' point and here the CFT is a $1+1$ sigma model \cite{orbifold2}. Below we describe this theory in the Euclidean signature. The 2D CFT lives on a base space spanned by a cylinder with coordinate
\bea
w= \tau + i\sigma
\label{w}
\eea 
where
\bea
-\infty < \tau <\infty,~~~~~0\leq \sigma <2\pi
\label{tau}
\eea

The target space is a symmetric product of $N_1N_5$ copies of $T^4$ under the action of the permutation group, $S_{N_1N_5}$ namely, 
\be
(T^4)^{N_1N_5}/S_{N_1N_5}
\ee
There are $4$ bosonic degrees of freedom $X^1,X^2,X^3,X^4$ and four fermionic degrees of freedom $\psi^1,\psi^2,\psi^3,\psi^4$ in the left moving sector and $4$ bosonic degrees of freedom $\bar X^1,\bar X^2,\bar X^3,\bar X^4$ and four fermionic degrees of freedom $\bar \psi^1,\bar \psi^2,\bar \psi^3,\bar \psi^4$ in the right moving sector. If the fermions have periodic boundary conditions on $S^1$ they are in Ramond $(R)$ sector and if they have antiperiodic boundary conditions on $S^1$ they are in the Neveu-Schwarz $NS$ sector. The central charge of the theory with fields $X^i,\psi^i, i =1 \ldots 4$ is $c=6$ and the total central charge with all the copies included is $ c= 6N_1N_5 $.

In this $S^N$ orbifold theory, there are twist sectors where $k$ copies of the CFT are joined together to give a single copy of the CFT on a circle of length $2\pi k$. We call each such set of joined copies a `component string' or `strand'.

\subsection{Symmetries of the CFT}

The D1D5 CFT has $(4,4)$ supersymmetry which means it has an  $\mathcal{N}=4$ left-moving supersymmetry and an $\mathcal{N}=4$ right-moving supersymmetry. Each algebra has an $R$ symmetry group which acts as a global symmetry that can be written as $SU(2)_L\times SU(2)_R$. 
Geometrically, in the CFT, this symmetry arises from an $SO(4)_E\simeq SU(2)_L\times  SU(2)_R$ symmetry which acts as rotations along the four spatial components of $M_{4,1}$. Here the subscript $E$ stands for external to label rotations along noncompact directions which correspond to $S^3$. The $T^4$ also has a symmetry $SO(4)_I\simeq SU(2)_1\times SU(2)_2$. The $I$ stands for internal to label rotations along internal directions. This symmetry is broken when the torus is compactified but at the orbifold point it still gives a useful organizing principle for the fields. Spinor indices  $\a,\bar{\a}$ are used for $SU(2)_L$ and $SU(2)_R$ respectively. Spinor indices $A,\dot A$ are used for $SU(2)_1$ and $SU(2)_2$ respectively. 
The four real fermions in the left moving sector can be grouped into complex fermions, $\psi^{\a A}$ and similarly the right moving fermions into $\bar{\psi}^{\bar \a A}$. The bosons $X^i$ transform as a vector on the $T^4$ but can be decomposed into complex scalars $X_{A\dot A}$ in the $({1\over2},{1\over2})$ representation of $SU(2)_1\times SU(2)_2$.

The left moving bosonic and fermionic excitations on a component string labeled by $i$ and with winding $k_{i}$ have the following mode expansions
\bea\label{modes}
\alpha^{(i)}_{A \dot A,m}&=&\frac{1}{2\pi }\int_{\sigma=0}^{2\pi k_{i}}\p_{w}X_{A \dot A}(w)e^{mw}dw~~~~~~m=\frac{q}{k_{i}}\nn
d^{\alpha A (i)}_{m}&=&\frac{1}{2\pi i }\int_{\sigma=0}^{2\pi k_{i}}\psi^{\alpha A}(w)e^{mw}dw
~~~~~~m=\frac{q}{k_{i}} ~~(\text{R}), ~~~m=\frac{q+\tfrac{1}{2}}{k_{i}} ~~(\text{NS})
\eea
Here $q$ is an integer. The R and NS denote the R sector and NS sector respectively. There are also the corresponding bosonic mode $\bar \alpha_{A\dot A,m}$ and fermionic mode $\bar d^{\bar\alpha A}_{m}$ for the right movers.

\subsection{Deformation of the CFT}

The orbifold CFT describes the system at its `free' point in moduli space. Since we are interested in a dynamical process, namely how a probe is tidally excited as it propagates within a superstratum geometry, we need to study a dynamical process in the CFT. This requires us to add an interaction to the theory.  To do this we deform the orbifold CFT by adding a marginal deformation, $D$, to the free action \cite{Avery:2010er,Burrington:2014yia,Carson:2016uwf,peet}
\be\label{defor S}
S\r S+\lambda \int d^2 z D(z, \bar z)
\ee
where $D$ has conformal dimensions $(h, \bar h)=(1,1)$. We choose the following form of $D$ which is a singlet under all symmetries at the orbifold point. It is given by
\be\label{D 1/4}
D=\epsilon^{\dot A\dot B}G^{-}_{\dot A, -\h} \bar G^{-}_{\dot B, -\h} \sigma^{++}
\ee
where $\sigma^{++}$ is a twist operator of rank $2$ in the theory. It can take two strands of the CFT of winding $k_1$ and $k_2$ and join them together into a strand of winding $k_1+k_2$. It can also take a strand of winding $k_1+k_2$ and split it into two strands of winding $k_1$ and $k_2$ respectively.

The twist operator in (\ref{D 1/4}) is defined as
\bea
\sigma = \sum_{i< j}\sigma_{ij}
\eea
where $\sigma_{ij}$ twists the two copies labelled by $i$ and $j$.
The operators $ G$ and $\bar G$ are the left and right moving supercharge operators at the orbifold point. 

In this paper we will study the large $N$ limit, where $N=N_1 N_5$.
According to \cite{Gomis:2002qi,gn}, 
we define $g$ as
\footnote{By matching the string spectrum in the PP-wave limit \cite{gn}, the coupling $\lambda$ in (\ref{defor S}) can be identified with the six-dimensional string coupling $g_6=g_s\sqrt{Q_5/Q_1}$. The radius of $AdS_3$ and $S^3$ is $(R_{AdS}/l_s)^2=g_6 \sqrt{N}$. Thus the first inequality in (\ref{gravity point}) arises from the requirement that the AdS radius is much larger than the string length ($R_{AdS}/l_s\gg 1$) while the second follows from the requirement that the string coupling itself is small ($g_s\ll 1$).}
\bea\label{t coupling}
g\equiv \lambda\,N^{1/2}  
\eea
where the coupling $g$ plays the role of the 't Hooft coupling. This coupling should not be confused with the string coupling $g_s$.
In the case where $N_1\sim N_5$, the perturbative CFT is described by $g\ll 1$, while 
the D1D5 supergravity solution is described by
the parameter region
\be\label{gravity point}
1\ll g \ll \sqrt{N}
\ee
In the following, we will do the computation in the perturbative CFT region where $g\ll 1$ and then extrapolate the results to the gravity region where $g$ is large.

\section{The splitting Process}\label{sec 3}

In this section we will discuss the process we will be looking at. One left and one right mover splits into three left and three right movers in the background of a $(1,0,n)$ superstratum state. In subsection \ref{ss state} we will introduce the $(1,0,n)$ superstratum state in the CFT. In subsections \ref{initial state} and \ref{final state} we will introduce the initial state and the final state of the splitting process respectively.

\subsection{CFT Dual of $(1,0,n)$ Superstratum}\label{ss state}

Superstata are smooth, backreacted, three charged 1/8-BPS solutions in supergravity. They correspond to states within the CFT with left-moving momentum excitations in the Ramond sector. 
For more details about superstrata and their CFT duals, see a recent review \cite{masaki}.

For the $(1,0,n)$ superstratum geometry, the dual CFT state is given by
\bea
|\psi\rangle &=& \bigg((L_{-1}-J^3_{-1})^n|00\rangle\bigg)^{N_{00}}\bigg(|++\rangle\bigg)^{N_{++}}
\label{ss state}
\eea
where 
\bea
|00\rangle &=&{1\over\sqrt2}\e_{AB}|AB\rangle = {1\over\sqrt2}\e_{AB} d^{-A}_0 \bar d^{-B}_0 |++\rangle \cr
|++\rangle &=&|0^+_R\rangle |\bar0^+_R\rangle 
\eea
with $|00\rangle ,|++\rangle $ composed of singly wound component strings.
Each of the $00$ strands with state $(L_{-1}-J^3_{-1})^n|00\rangle$ correspond to a $1/8$-BPS supergraviton.
We'll call $|\psi\rangle$  the $(1,0,n)$ state. Unbarred states and operators denote left moving quantities and barred states and operators denote right moving quantities.
The $++$ strands correspond to Ramond states charged under $SU(2)_L\times SU(2)_R$ with quantum numbers
\bea
j_L=j_R={1\over2}
\eea
which are proportional to the angular momentum along the directions of $S^3$. $N_{++}$ counts the number of these strands.
Here $N_{00}$  is the number of $00$ strands.
The total winding number is given by
\bea
N\equiv N_1N_5 = N_{00} + N_{++}
\eea
Below we record the precise relationship between the parameters in the $(1,0,n)$ superstratum geometry and the CFT winding numbers:
\bea
N_{++} = \mathcal{N}a^2\cr
N_{00} =  \mathcal{N}{b^2\over2}
\eea
where
\bea
\mathcal{N} = {V_4R_y^2\over(2\pi)^4g_s^2\a'^4}
\eea
Here $V_4$ is the volume of $T^4$, $g_s$ is the $10D$ string coupling.
While the state \ref{ss state} will be the focus of this paper there are, however, other superstrata geometries with CFT duals which are also $1/8$-BPS. These are obtained by acting with the appropriate number of $J^+_{-1}$'s and $G^{+}_{\dot A, -1}$'s. For more details see \cite{masaki}.

\subsection{The Initial State}\label{initial state}

In \cite{mw,chl} the authors considered a radially infalling graviton in a $(1,0,n)$ superstratum geometry and showed that the kinetic energy of the infalling probe was converted to string excitations. 
We are interested in the CFT dual of a graviton propagating within the superstratum geometry. 

To start we take the $(1,0,n)$ state, (\ref{ss state}), and replace one of the vaccum $++$ strands by a probe graviton which corresponds to a $++$ strand carrying one left and one right moving boson. The state is
\be\label{initial N}
\bigg((L_{-1}-J^3_{-1})^n|00\rangle\bigg)^{N_{00}}\bigg(|++\rangle\bigg)^{N_{++}-1}\a_{++,-m}\bar\a_{--,-m} |++\rangle
\ee
which is depicted in fig.\,\ref{fig_states}.

Let us first take the probe graviton and one of the $00$ strands from (\ref{initial N})
\be
(L^{(i)}_{-1}-J^{3(i)}_{-1})^n|00\rangle^{(i)} \a^{(j)}_{++,-m}\bar\a^{(j)}_{--,-m}|++\rangle^{(j)}
\ee
where $i,j=1,2$ are copy labels.
In section \ref{sec 6} we will include other strands and consider the large $N$ limit. 
Since we are only working with two singly wound copies of the CFT we have included explicit copy indices as superscripts to label the copies.

One can show that 
\bea
(L^{(i)}_{-1} - J^{3(i)}_{-1})^n|00\rangle^{(i)} = {n!\over\sqrt2}\e_{AB}d^{-A(i)}_{-n}\bar d^{-B(i)}_0| ++\rangle^{(i)}
\eea
where $d$ and $\bar d$ refer to fermionic excitations in the CFT. 
Define
\bea
|\psi_{i,j}\rangle &=& (L^{(i)}_{-1}-J^{3(i)}_{-1})^n|00\rangle^{(i)}|++\rangle^{(j)}\nn
&=& {n!\over\sqrt2}\e_{AB}d^{-A(i)}_{-n}\bar d^{-B(i)}_0|++\rangle^{(i)} |++\rangle^{(j)}
\eea
We therefore write our initial state in the following way
\bea
|\Psi_0\rangle &\equiv&{1\over \sqrt2mn!}\a^{(2)}_{++,-m}\bar\a^{(2)}_{--,-m} |\psi_{1,2}\rangle + {1\over \sqrt2mn!}\a^{(1)}_{++,-m}\bar\a^{(1)}_{--,-m} |\psi_{2,1}\rangle\cr
&=& {1\over2m}\a^{(2)}_{++,-m}\bar\a^{(2)}_{--,-m} \e_{AB}d^{-A(1)}_{-n}\bar d^{-B(1)}_0|++\rangle^{(1)}|++\rangle^{(2)}  \cr
&& +  {1\over2m}\a^{(1)}_{++,-m}\bar\a^{(1)}_{--,-m} \e_{AB}d^{-A(2)}_{-n}\bar d^{-B(2)}_0|++\rangle^{(2)}|++\rangle^{(1)}
\label{Psi}
\eea
Since these are states built from the symmetric product orbifold, we have symmetrized over the two singly wound copies. We also note that $|\Psi_0\rangle$ has been normalized to $1$.

\subsection{Final State}\label{final state}

The gravity computation in \cite{mw,chl} showed that the kinetic energy of a radially infalling string probe was converted into stringy excitations. Thus in the CFT we will consider a final state where the strand corresponding to a probe graviton is converted to a strand corresponding to a single particle stringy state. As discussed in \cite{Guo:2021ybz}, the simplest final state is
\be\label{final N}
\bigg((L_{-1}-J^3_{-1})^n|00\rangle\bigg)^{N_{00}}\bigg(|++\rangle\bigg)^{N_{++}-1}
\Big(\alpha_{++,-p} \,d^{-+}_{-q}\,d^{+-}_{-r} \,
\bar\alpha_{--,-{p}}\, \bar d^{+-}_{-q}\,\bar d^{-+}_{-r}
|++\rangle\Big)
\ee
which is depicted in fig.\,\ref{fig_states}. Compared to the initial state (\ref{initial N}), the probe strand with one left and one right mover is converted to a strand with three left and three right movers.
The resulting strand is generally believed to correspond to a single particle stringy state since the excitations cannot be written only in terms of the anomaly-free part of the generators of the symmetry.
Other strands are unchanged. This corresponds to a fixed superstratum background in gravity. For recent progress in understanding the tensionless string in $AdS_3$ from the CFT at the orbifold point, see \cite{Eberhardt}.

Taking the strand corresponding to a stringy state along with one of the $00$ strands, we have the final state
\bea
|\Psi_f\rangle &\equiv&{1 \over \sqrt2pn!}\alpha^{(2)}_{++,-p} d^{-+(2)}_{-q}d^{+-(2)}_{-r} 
\bar\alpha^{(2)}_{--,-{p}} \bar d^{+-(2)}_{-q}\bar d^{-+(2)}_{-r}
|\psi_{1,2}\rangle\cr
&&+{1 \over \sqrt2pn!}\alpha^{(1)}_{++,-p} d^{-+(1)}_{-q}d^{+-(1)}_{-r} 
\bar\alpha^{(1)}_{--,-{p}} \bar d^{+-(1)}_{-q}\bar d^{-+(1)}_{-r}
|\psi_{2,1}\rangle
\label{Psi f}
\eea
The conjugate is 
\bea
\langle\Psi_f| &\equiv& {1\over \sqrt2pn!}\langle\psi_{1,2}| \alpha^{(2)}_{--,p} d^{+-(2)}_{q}d^{-+(2)}_{r} \bar\alpha^{(2)}_{++,{p}} \bar d^{-+(2)}_{q}\bar d^{+-(2)}_{r}\cr
&&+{1\over \sqrt2pn!}\langle\psi_{2,1}| \alpha^{(1)}_{--,p} d^{+-(1)}_{q}d^{-+,(1)}_{r} \bar\alpha^{(1)}_{++,{p}} \bar d^{-+(1)}_{q}\bar d^{+-(1)}_{r}
\cr
&=&{1\over2p}\e_{BA}{}^{(1)} \langle ++| {}^{(2)}\langle++| d^{+A(1)}_{n}\bar d^{+B(1)}_0 \alpha^{(2)}_{--,p} d^{+-(2)}_{q}d^{-+(2)}_{r} \bar\alpha^{(2)}_{++,{p}} \bar d^{-+(2)}_{q}\bar d^{+-(2)}_{r} \cr
&&+{1\over2p}\e_{BA}{}^{(2)} \langle ++| {}^{(1)}\langle++| d^{+A(2)}_{n}\bar d^{+B(2)}_0 \alpha^{(1)}_{--,p} d^{+-(1)}_{q}d^{-+(1)}_{r} \bar\alpha^{(1)}_{++,{p}} \bar d^{-+(1)}_{q}\bar d^{+-(1)}_{r} 
\label{Phi}
\nn
\eea
The dimensions of the final state are
\bea
h = p + q + r + n, ~~~\bar h = p + q + r
\eea
We note that we have written the form of the final state as a \textit{bra} which is convenient for our context since we'll be computing a transition amplitude. 
The final state is normalized as $\langle\Psi_f| \Psi_f\rangle=1$. Now that we have written our initial and final states let us write down the transition amplitude we would like to compute.
\section{The Amplitude}\label{sec 4}
In this section we discuss the method for computing the amplitude. Let us recall the process we are interested in. We are concerned with how one left and one right mover in the initial state, (\ref{Psi}), splits into three left and three right movers in the final state, (\ref{Phi}) using the deformation operator, $D$ given in (\ref{D 1/4}).
\be
|\Psi_0\rangle \rightarrow |\Psi_f\rangle
\ee
In this process, we need the application of two deformation operators, $DD$, where the first $D$ joins two singly wound component strings into a doubly wound component string and the second $D$ untwists the doubly wound string back into two singly wound component strings. 
Using (\ref{Psi}), (\ref{defor S}), (\ref{D 1/4}) and (\ref{Phi}), the full amplitude we would like to compute, which is at second order in the coupling $\lambda$, is given  by 
\bea
A_{m,n}^{0 \to f}(\tau)&\equiv&{1\over2}\lambda^2\int d^2w_2d^2w_1 \langle\Psi_f| D(w_2,\bar w_2) D(w_1,\bar w_1) |\Psi_0\rangle
\cr
\cr
&\equiv&{1\over2}\lambda^2\int d^2w_2d^2w_1\mathcal{A}^{0 \to f}(w_1,w_2,\bar w_1, \bar w_2)
\label{amplitude}
\eea
with the locations of the twists given by
\bea
w_i &=& \tau_i + i\sigma_i\cr
\bar{w}_i &=& \tau_i - i \sigma_i
\eea 
The integration region is taken to be 
\bea
-{\tau\over2}\leq \tau_i \leq {\tau\over2},\quad 0\leq \sigma_i< 2\pi
\eea
The expression for the amplitude $\mathcal{A}$ is given by
\vspace{1mm}
\bea\label{amplitude 1}
&&\!\!\!\mathcal{A}^{0 \to f}(w_1,w_2,\bar w_1,\bar w_2)\nn[1.5pt]
&&=  \epsilon^{\dot A \dot B}\epsilon^{\dot C \dot D}\langle  \Psi_f | G^+_{\dot C,-{1\over2}}\s^-_2(w_2)G^-_{\dot A,-{1\over2}}\s^+_2(w_1) \bar G^+_{\dot D,-{1\over2}}\bar \s^-_2(\bar w_2)\bar G^-_{\dot B,-{1\over2}}\bar\s^+_2(\bar w_1)|\Psi_0\rangle
\eea
In the next section we will outline the steps to compute these amplitudes without going through all the details. They are involved but straightforward. 

\subsection{The Computation}

Here we outline the computation of the amplitude, (\ref{amplitude 1}). The techniques for computing these amplitudes were developed over the course of a series of papers \cite{Carson:2016uwf}. Some of the expressions used here have been exactly computed in previous works while others, though not computed explicitly previously, follow from the same techniques. It is not so helpful in presenting them in full detail as they would detract from the main physics. Let us begin.
\begin{enumerate}
\item The amplitude, (\ref{amplitude 1}), is defined on a base space which is a cylinder defined by the coordinates (\ref{w}) and (\ref{tau}). 
\item The presence of the twist operators change the boundary conditions of the fields on the cylinder making them double-valued. This renders computations ill-defined. Therefore, to alleviate this problem we must map to a covering space. This is first done by mapping to the complex plane through the map
\bea
z=e^w
\eea
and then to the covering space which is given by the map
\bea
z = {(t+a)(t+b)\over t}
\eea
For more details on the covering map see \cite{Carson:2016uwf}. 
\item When mapping to the $t$-plane the amplitude can be written in the following schematic form
\bea
\mathcal A \sim C\mathcal A_t
\eea
The constant $C$ comes from various Jacobian factors which arise from coordinate transformations of fermionic and bosonic modes from the cylinder to the $t$ plane. The details of these transformations are given \cite{dissertation}. There are also other operator insertions which are present. The amplitude $\mathcal{A}_t$ contains wick contraction terms between pairs of fermionic and bosonic modes which have undergone all necessary transformations. The details for computing such terms are given in \cite{Carson:2016uwf}. While not every term which was computed for this paper is given explicitly in \cite{dissertation}, the method for computing such terms are the same as those given there.
\item Finally, we note that there are both holomorphic and antiholomorphic components to the amplitude. These components are multiplied together to give (\ref{amplitude 1}).
\end{enumerate}

After carrying out the above steps we find that the amplitude can be written in the following form 
\bea
\mathcal{A}^{0 \to f}(w_1,w_2,\bar w_1,\bar w_2) &=&\sum_{k=k_{min}(m,n,p,q,r)}^{k_{max}(m,n,p,q,r)}\sum_{\bar k = \bar k _{min}(m,n,p,q,r)}^{\bar k_{max}(m,n,p,q,r)}B^{0\to f}_{k,\bar k}(m,n,p,q,r)e^{{k\Delta w\over2} + {\bar k\Delta \bar w\over2}}\nn
\eea
This is the same form we find in \cite{hm,dissertation,Guo:2021ybz}. We find the following limits for the sums
\bea
k_{min}(m,n,p,q,r) &=& -2(m+n) \cr
k_{max}(m,n,p,q,r) &=& 2(m+n)\cr
\bar k_{min}(m,n,p,q,r) &=& -2m \cr
\bar k_{max}(m,n,p,q,r)&=& 2m
\eea
The coefficients $B^{0\to f}_{k,\bar k}$ are nonzero when $k-\bar k$ is an even integer. The diagonal elements $k=\bar k$ have the symmetry
\be\label{B symmetry}
B^{0 \to f}_{k,k}(m,n,p,q,r) = B^{0\to f}_{- k,- k}(m,n,p,q,r)
\ee

\subsection{Integrating the Amplitude}

In this section we integrate over the twist insertions to obtain the full amplitude. We begin with the expression
\bea
A^{0\to f}_{m,n}(\tau)&=&{1\over2}\lambda^2\int d^2w_2d^2w_1\mathcal{A}^{0 \to f}(w_1,w_2,\bar w_1,\bar w_2)\cr
&=&{1\over2}\lambda^2\int d^2w_2d^2w_1\sum_{k=k_{min}(m,n,p,q,r)}^{k_{max}(m,n,p,q,r)}~~\sum_{\bar k = \bar k _{min}(m,n,p,q,r)}^{\bar k_{max}(m,n,p,q,r)}B^{0\to f}_{k,\bar k}(m,n,p,q,r)e^{{k \Delta w\over2}+{\bar k\Delta \bar w\over2}}\cr
&=&\lambda^2\sum_{k=k_{min}(m,n,p,q,r)}^{k_{max}(m,n,p,q,r)}~~\sum_{\bar k = \bar k _{min}(m,n,p,q,r)}^{\bar k_{max}(m,n,p,q,r)}B^{0 \to f}_{k,\bar k}(m,n,p,q,r)I_{k,\bar k}
\eea
where
\bea
I_{k,\bar k} \equiv{1\over2} \int d^2w_2d^2w_1e^{{k \Delta w\over2}+{\bar k \Delta \bar w\over2}}
\eea
The twist separations in the left(right) moving sectors are given by
\bea
\Delta w &=& w_2 - w_1=\tau_2 - \tau_1 + i(\sigma_2 - \sigma_1)\cr
\Delta \bar w &=& \bar w_2 - \bar w_1 =\tau_2 - \tau_1 - i(\sigma_2 - \sigma_1)
\eea
The $B^{0\to f}_{k,\bar k}(m,n,p,q,r)$'s are numerical coefficients which depend on the initial energy $m$, the superstratum momentum $n$ and the final energies $p,q,r$. To obtain a real space amplitude we must wick rotate back to Lorentzian signature by taking $\tau \to it$. This gives
\bea
\Delta w &=& w_2 - w_1=i(t_2 - t_1 + \sigma_2 - \sigma_1)\cr
\Delta \bar w &=& \bar w_2 - \bar w_1 =i(t_2 - t_1 - (\sigma_2 - \sigma_1))
\eea
The integration region is now 
\bea
-{ t\over2}\leq t_i \leq {t\over2},~~~~\quad 0\leq \sigma_i< 2\pi
\eea
Our expression for $A^{0\to f}_{m,n}$ becomes
\bea
A^{0\to f}_{m,n}(t)&=&\lambda^2\sum_{k=k_{min}(m,n,p,q,r)}^{k_{max}(m,n,p,q,r)}\sum_{\bar k = \bar k _{min}(m,n,p,q,r)}^{\bar k_{max}(m,n,p,q,r)}B^{0\to f}_{k,\bar k}(m,n,p,q,r)I_{k,\bar k}(t)
\label{amplitude}
\eea
where
\bea
I_{k,\bar k}(t)&=&\frac{1}{2}\int d^2w_2\,d^2w_1\, e^{{k \Delta w\over2}+{\bar k \Delta \bar w\over2}}\cr
&=&\int_{-{t\over2}}^{{t\over2}}dt_2\int_{-{t\over2}}^{t_2}dt_1\int_{\s_2=0}^{2\pi} d\s_2\int_{\s_1=0}^{2\pi} d\s_1 e^{{i(k+\bar k)\over2}(t_2-t_1)}e^{{i(k-\bar k)\over2}(\s_2-\s_1)}
\label{integral}
\eea
Here, $t$ is duration of the two deformation operators. 
Notice that since $k-\bar k$ is an even integer, $\frac{k-\bar k}{2}$ is an integer. The $\sigma_i$ integrals enforce $k=\bar k$, namely momentum conservation in the intermediate state. We also enforce energy conservation between initial and final states
\bea
n + m = n + p + q + r
\eea
There are two contributions to the integration depending on whether $k=\bar k$ is zero or nonzero. The first contribution, being
\bea
k=\bar k \neq 0
\eea
gives
\bea
I_{k=\bar k\neq 0}(t)
&=&\frac{1}{2}\int d^2w_2\,d^2w_1\, e^{{k \Delta w\over2}+{\bar k \Delta \bar w\over2}}\cr
&=&\int_{-{t\over2}}^{{t\over2}}dt_2\int_{-{t\over2}}^{t_2}dt_1\int_{\s=0}^{2\pi}d\s_2\int_{\s=0}^{2\pi}d\s_1\, e^{i k (t_2-t_1)}\cr
&=& {4i \pi^2 \over m ^2} \bigg(kt  -  2e^{i {kt\over2}}\sin\bigg(  {kt\over 2}\bigg)  \bigg)
\label{integral three}
\eea
The second contribution being
\bea
k=\bar k = 0
\eea
gives
\bea
I_{k=\bar k= 0}(t) &=&\frac{1}{2}\int d^2w_2\,d^2w_1\, e^{{k \Delta w\over2}+{\bar k \Delta \bar w\over2}}\cr
&=&\int_{-{t\over2}}^{{t\over2}}dt_2\int_{-{t\over2}}^{t_2}dt_1\int_{\s=0}^{2\pi}d\s_2\int_{\s=0}^{2\pi}d\s_1\cr
&=&2\pi^2t^2
\eea
In this computation the $k$ and $\bar k$ sums don't have the same ranges since the left and right moving amplitudes differ. Recall that the superstratum state only has left moving momentum. Therefore in the case where $k = \bar k$ and nonzero, we sum over $\bar k$ because it always has the smallest range.

Recall that the coefficients have the following symmetry (\ref{B symmetry})
\bea
B^{0 \to f}_{ k, k}(m,n,p,q,r) = B^{0\to f}_{- k,- k}(m,n,p,q,r)
\eea
Therefore the terms linear in $t$ from the contribution $k=\bar k \neq 0$ (\ref{integral three}) are cancelled.
We can write the full answer for the amplitude as the following
\bea
A^{0 \to f}_{m,n}(t)&=&\lambda^2 2\pi^2t^2B^{0 \to f}_{0,0}(m,n,p,q,r)  + \lambda^2\sum_{  k = 1}^{ 2m }B^{i \to j}_{ k, k}(m,n,p,q,r){16\pi^2\over k^2}\sin^2\bigg({ kt\over2}\bigg)\nn
&\equiv&A^{{0 \to f},scr}_{m,n}(t) + A^{ {0 \to f},osc}_{m,n}(t)
\label{int amplitude}
\eea 
The oscillatory term in the amplitude, $A^{ {0 \to f},osc}_{m,n}(t)$, can be interpreted \cite{hm,Guo:2021ybz} as the graviton propagating within the geometry starting from the boundary. If the geometry was just $AdS_3\times S^3\times T^4$ as is the case of NS vacuum studied in \cite{hm,Guo:2021ybz}, the graviton would simply propagate from one side of the geometry to the other unhindered and then back again.

The $t^2$ term, $A^{{0 \to f},scr}_{m,n}(t)$, is the secular term responsible for tidal excitations where the superscript `\textit{scr}' stands for `scrambling'.  The $t^2$ growth indicates that this mode is encountering something, the superstratum geometry, which is converting it's energy into stringy excitations.  Furthermore, as $t^2$ grows, it becomes the dominant term overtaking the oscillatory term indicating that in the gravity dual the graviton never makes it back to the boundary but becomes trapped as is shown in \cite{mw}. 

In the next section we will compute the amplitude numerically and particularly focus on the $t^2$ term since this is the term which signals the presence of tidal excitations.

\section{Numerical Results}\label{sec 5}

In this section we numerically compute the amplitude in (\ref{int amplitude}) for various values of initial energy, $m$, superstratum momentum, $n$, and final energies $p=q=r=\frac{m}{3}$ with energy conservation enforced. 
By extrapolating the numerical results, we obtain an analytic expression of the amplitude for large $m$, which corresponds to a large radial kinetic energy of the infalling graviton.

\subsection{Computing $A^{0\to f}_{m,n}(t)$}

We numerically compute and plot the full amplitude $A^{0\to f}_{m,n}(t)$ given in (\ref{int amplitude}), as a function of $t$ for various values of initial probe energy $m$ and $n=1$. We consider the equal splitting case where $p=q=r={m\over3}$. 
\begin{figure}
\centering
        \includegraphics[width=7.85cm]{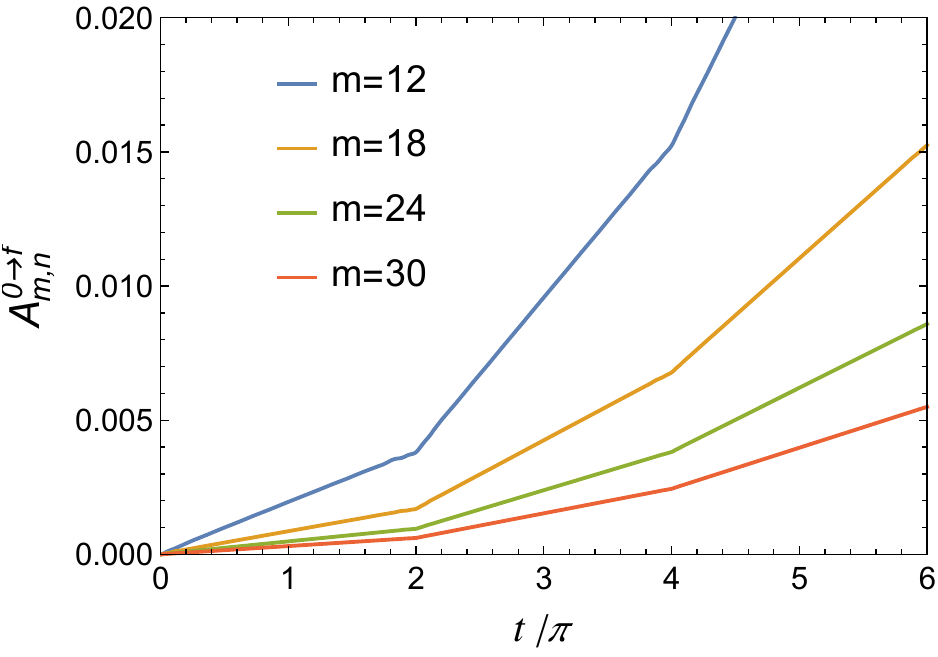}\quad
       \includegraphics[width=7.3cm]{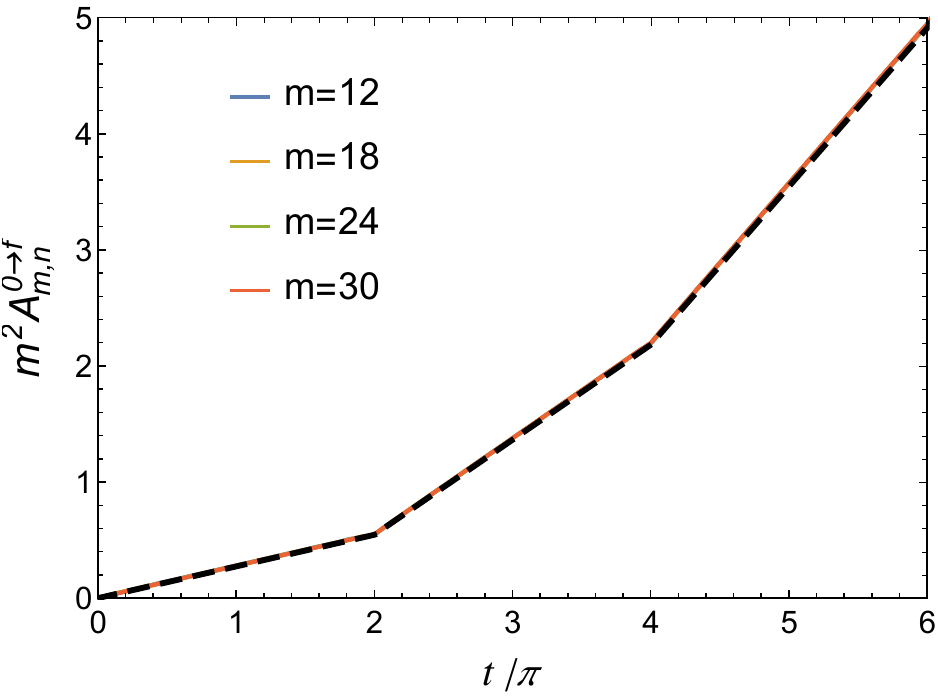}
\caption{Left: $A^{0\to f}_{m,n=1}(t)$ for $n=1$ and different $m$. Right: $m^2 A^{0\to f}_{m,n=1}(t)$ for different $m$ and $C \Big( 2\pi^2t^2  + \left[-2 \pi^2 t^2 +4\pi^3t\right]_{\text{saw-like},2\pi}\Big)$ with $C=0.0007$. In the plot, we set $\lambda^2=1$.}
\label{fig_amp}
\end{figure}
The amplitude $A^{0\to f}_{m,n=1}(t)$ and the rescaled amplitude $m^2A^{0\to f}_{m,n=1}(t)$ as a function of $t$ for different $m$ are shown in fig.\,\ref{fig_amp}. In the right panel, the black dashed line is $C \Big( 2\pi^2t^2  + \left[-2 \pi^2 t^2 +4\pi^3t\right]_{\text{saw-like},2\pi}\Big)$ with $C=0.0007$. All the lines in the right panel coincide. 
Here `saw-like, $2\pi$' means that we start with the function in the square bracket in the region $(0,\pi)$ and then reflect it across $\pi$ to obtain the function in the region $(\pi,2\pi)$. Then we make a saw-like function with periodicity $2\pi$.

Thus we find that for $n=1$
\bea\label{amp analytic}
A^{0\to f}_{m,n}(t)\approx \frac{C}{m^2}\lambda^2\Big( 2\pi^2t^2  + \left[-2 \pi^2 t^2 +4\pi^3t\right]_{\text{saw-like},2\pi}
\Big)
\label{int amplitude 0}
\eea
where
\bea
A^{0\to f,scr}_{m,n}(t)\approx  \frac{C}{m^2}\lambda^2 2\pi^2t^2,\qquad  A^{0\to f,osc}_{m,n}(t)\approx  \frac{C}{m^2}\lambda^2 \left[-2 \pi^2 t^2 +4\pi^3t\right]_{\text{saw-like},2\pi}
\eea
Actually this is correct for all $n$ as we will investigate in the next section. We indeed see that there is a growth of the amplitude due to the $t^2$ term. Since the final state corresponds to a single particle string state moving in the superstratum geometry, the $t^2$ secular term corresponds to the tidal forces created by the superstratum geometry acting on the string state. There is no longer a periodicity within the amplitude. On the gravity side, this shows that the graviton doesn't return to its starting place on the boundary after a period of $2\pi$.

As $m$ becomes larger, the amplitude converges to (\ref{amp analytic}) quickly. This large $m$ limit corresponds to a large radially infalling kinetic energy of the infalling graviton.   

\subsection{Properties of $B^{0\to f}_{ k, k}$}\label{sec B}

To better understand the properties of the amplitude in the large $m$ limit, i.e., a large radially infalling kinetic energy limit, in this section we will study the properties of the coefficient $B^{0\to f}_{ k, k}$ in the amplitude (\ref{int amplitude}). 

\begin{figure}
\centering
        \includegraphics[width=5.1cm]{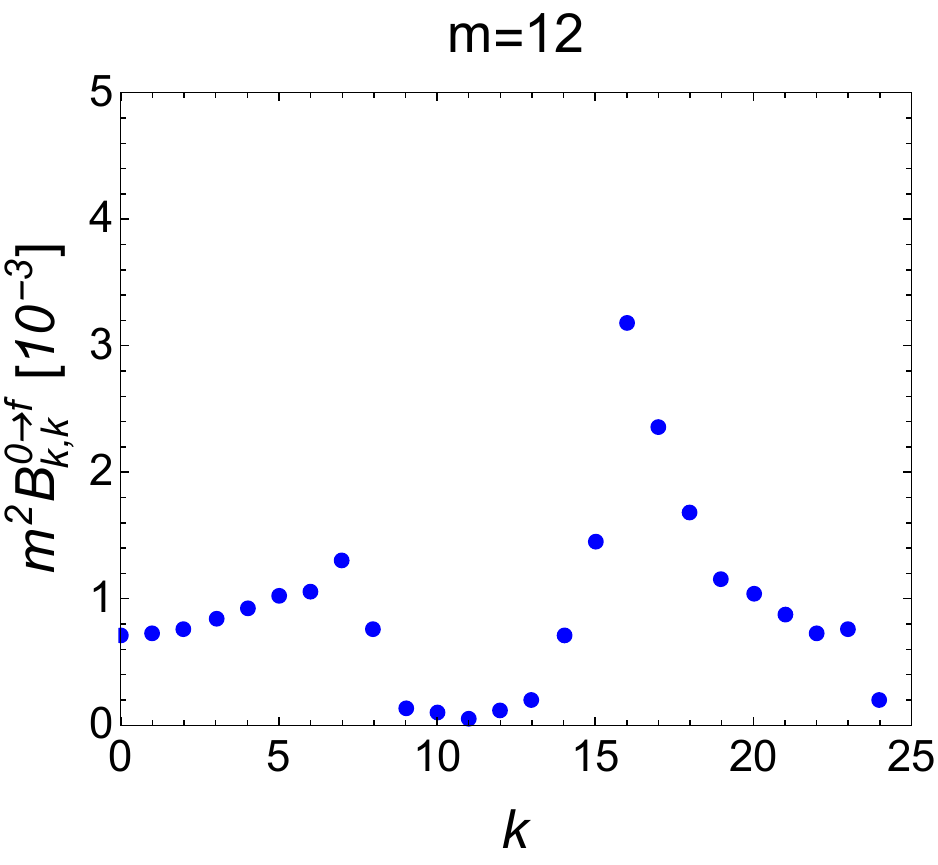}\quad
       \includegraphics[width=5cm]{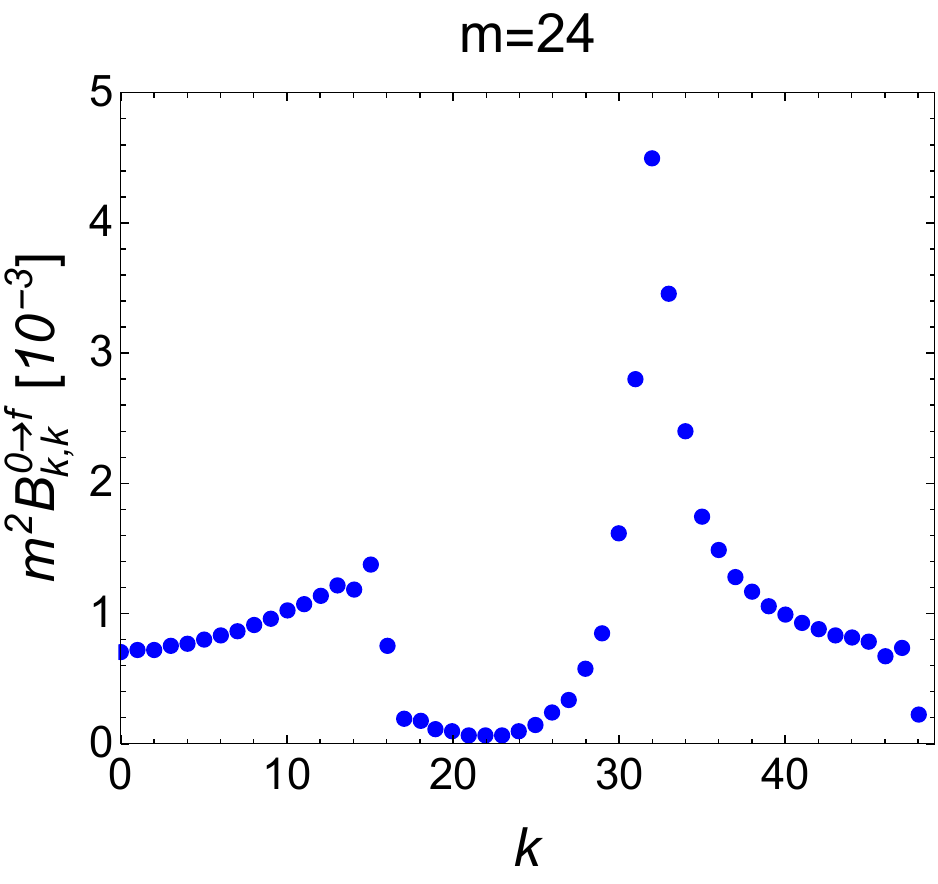}\quad
       \includegraphics[width=5cm]{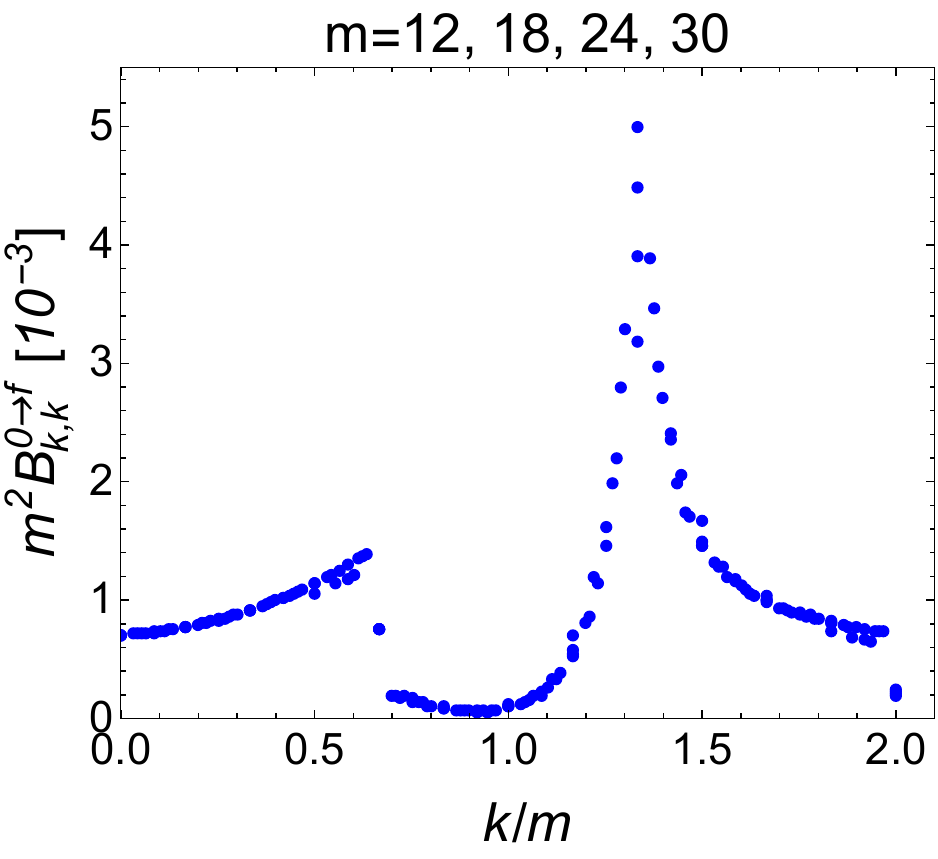}
\caption{$m^2 B^{0\to f}_{k, k}$ as a function of $m$ for $m=12$ (Left) and $m=24$ (Middle).
$m^2 B^{0\to f}_{k, k}$ as a function of $\frac{k}{m}$ for $m=12,18,24,30$ (Right).}
\label{fig_B}
\end{figure}

In fig.\,\ref{fig_B}, we plot $m^2 B^{0\to f}_{ k,  k}$ as a function of $k$ for different $m$ for $n=1$. We can see that for different $m$ the plots follow the same shape. As $m$ becomes larger, more points fit into this curve. We expect that this is true in the limit $m\rightarrow \infty$.

\begin{figure}
\centering
        \includegraphics[width=8cm]{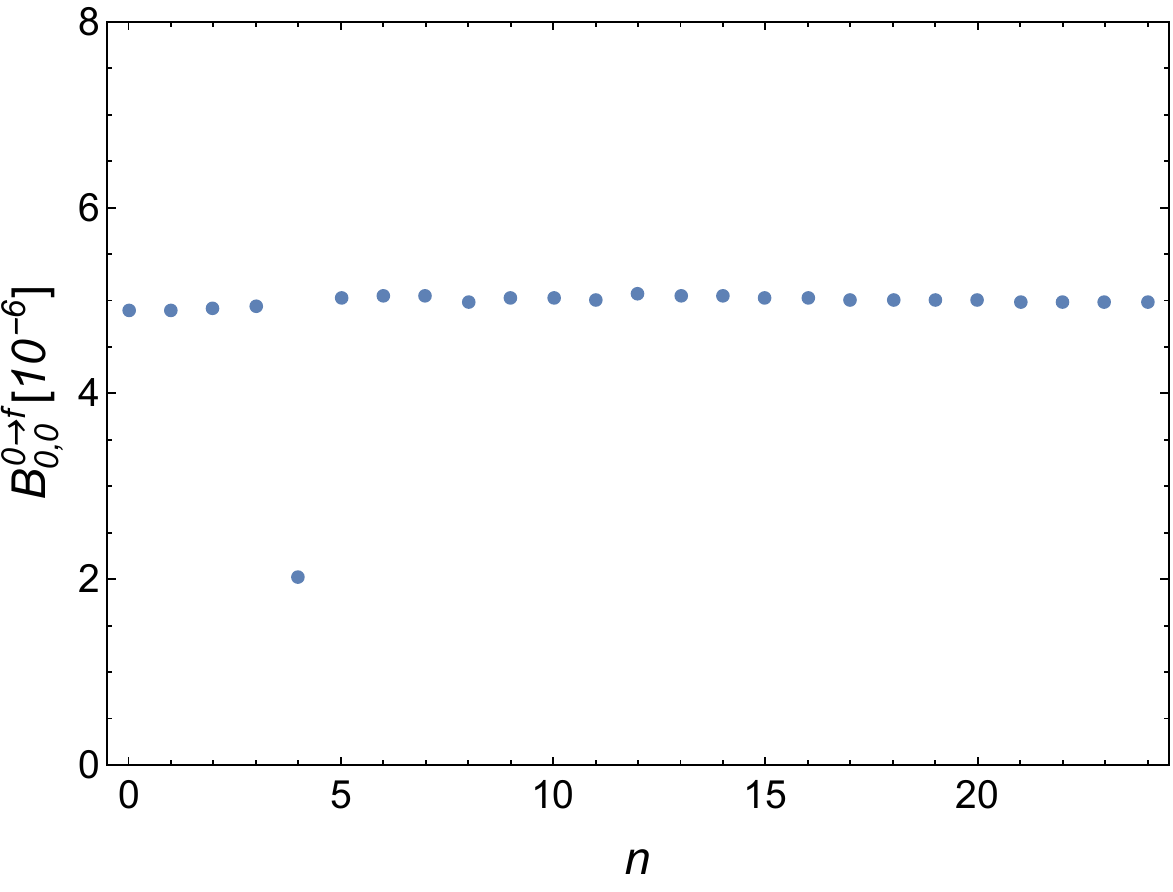}
\caption{$B^{0\to f}_{0,0}$ as a function of $n$ for $m=12$. It does not depend on $n$ except for when $n=m/3$.}
\label{fig_n}
\end{figure}

Thus as $m$ becomes larger, more and more points fit into the small $ k$ region, say the region $ k\lesssim m/6$. In this region $m^2 B^{0\to f}_{ k, k}$ can be approximated as a constant. From the right panel of fig.\,\ref{fig_B}, we find that the constant is
\be
C=m^2 B^{0\to f}_{ k, k}\approx m^2 B^{0\to f}_{0,0} \approx 0.0007~~~~~~~  k \ll m ~~~~~~m\rightarrow \infty
\ee
In fig.\,\ref{fig_n}, the $B^{0\to f}_{0,0}$ for $m=12$ and different $n$ are plotted. We see that $B^{0\to f}_{0,0}$ does not depend on $n$ except for when $n=q=m/3$ where $q$ is the mode number of one of the fermionic modes in the final state (\ref{Psi f}). In this case the value of $B^{0\to f}_{0,0}$ drops because of the Pauli Exclusion Principle. We think that a similar reason applies to fig.\,\ref{fig_B} where some of the points drop below the expected value. Thus the above constant $C$ does not depend on $n$. 
In this approximation, the amplitude $A^{0\to f}_{m,n}(t)$ (\ref{int amplitude}) becomes
\bea
A^{0\to f}_{m,n}(t)&\approx&\frac{C}{m^2}\lambda^2\Big[ 2\pi^2t^2  + \sum_{ k = 1}^{\infty}{16\pi^2\over k^2}\sin^2\bigg({ kt\over2}\bigg)\Big]
\label{int amplitude 1}
\eea 
The second oscillation term with periodicity $2\pi$ can be found by
\footnote{It is instructive to compare to
\be
\sum_{ k = 1\in\mathbb{Z}_{\text{odd}}}^{\infty}{1\over k^2}\sin^2\bigg({ kt\over2}\bigg) = \frac{\pi}{8} [t]_{\text{saw-like},2\pi}
\ee
which appears in the study of a free falling graviton in AdS \cite{Guo:2021ybz}.}
\be
\sum_{ k = 1}^{\infty}{1\over k^2}\sin^2\bigg({ kt\over2}\bigg) = \left[-\frac{1}{8}t^2 +\frac{\pi}{4}t\right]_{\text{saw-like},2\pi}
\ee
Thus in the large $m$ limit, we have
\bea
A^{0\to f}_{m,n}(t)= \frac{C}{m^2}\lambda^2\Big( 2\pi^2t^2  + \left[-2 \pi^2 t^2 +4\pi^3t\right]_{\text{saw-like},2\pi}
\Big)
\label{int amplitude 1}
\eea
We see that for $t$ in the region $(0,\pi)$ there is a cancellation of the $t^2$ term. Only the linear $t$ term is left and the amplitude looks like the one where the splitting of an initial probe was computed in the NS vacuum \cite{Guo:2021ybz} which describes a particle moving in empty AdS. 
In this paper, the linear $t$ behavior in the region $(0,\pi)$ corresponds to an infalling graviton becoming redshifted when it travels down the throat of the superstratum geometry. At late times, i.e., $t\gtrsim \pi$, the $t^2$ term will not be cancelled and this corresponds to the tidal forces created by the superstratum geometry acting on the string state.

The answer in (\ref{int amplitude 1}) can be rewritten as
\bea
A^{0\to f}_{m,n}(t)= \frac{C}{m^2}\lambda^2\bigcup_{n=0}^{n_0}\Big( (8n+4) \pi^3 t - 8n(n+1)\pi^4
\Big)_{t\in[n2\pi,(n+1)2\pi]}
\eea
where $n_0$ is sufficiently high enough at which point the perturbation theory breaks down.
This explains the fact that in each of the regions $t\in[n2\pi,(n+1)2\pi]$, the plot in fig.\,\ref{fig_amp} is linear.

\section{Large $N$ limit}\label{sec 6}

In this section we consider the large $N$ limit. 
The initial state (\ref{initial N}) corresponds to a graviton in the $(1,0,n)$ superstratum background
\bea\label{N initial}
|\Psi_{0};N\rangle &=& \bigg(\frac{1}{n!}(L_{-1}-J^3_{-1})^n|00\rangle\bigg)^{N_{00}}\bigg(|++\rangle\bigg)^{N_{++}-1}\Big({1 \over m}\,\a_{++,-m}\bar\a_{--,-m} |++\rangle\Big)
\eea
The two deformation operators can twist and untwist the strand corresponding to the graviton with any of the $00$ or $++$ strands. The twist and untwist with the $++$ strands has been studied in \cite{Guo:2021ybz}. It was found that the amplitude is periodic without any secular term.
To obtain a secular term, it has to twist and untwist with the $00$ strands. This represents the probe interacting with the superstratum geometry in the gravity dual.
After applying two deformation operators, the strand with one left and one right mover becomes a strand with three left and three right movers, which corresponds to a single particle stringy state.
The final state is
\bea
|\Psi_{f};N\rangle &\equiv& \bigg(\frac{1}{n!}(L_{-1}-J^3_{-1})^n|00\rangle\bigg)^{N_{00}}\bigg(|++\rangle\bigg)^{N_{++}-1}\nn
&&\Big({1 \over p}\,\alpha_{++,-p} \,d^{-+}_{-q}\,d^{+-}_{-r} \,
\bar\alpha_{--,-{p}}\, \bar d^{+-}_{-q}\,\bar d^{-+}_{-r}
|++\rangle\Big)
\label{Psi f}
\eea
The process is
\be
|\Psi_{0};N\rangle \rightarrow  |\Psi_{f};N\rangle
\ee
The amplitude for the case $N_{00}=N_{++}=1$ is $A^{0\to f}_{m,n}$.
In the following, let us focus on the secular terms, which are important at long time scales. 
Recall that the number of $00$ strands corresponding to supergravitons in the superstratum state is $N_{00}$.
Because the two deformation operators can twist and untwist with any of the $00$ strands, the amplitude is
\be
\tilde{A}^{scr}_{m,n}(t) = N_{00} A^{0\to f,scr}_{m,n} =  N_{00} \lambda^2 2\pi^2 {C\over m^2} t^2 = N_{00} \lambda^2 2\pi^2 {C\over m^2} t^2 =\frac{N_{00}}{N} g^2 2\pi^2 {C\over m^2} t^2
\ee
where we only include the growing term at large $t$, $A^{{0 \to f}, scr}_{m,n}(t)$, in (\ref{int amplitude 1}). Here we have used $\tilde{A}$ to represent the amplitude at large $N$.

When the number of supergravitons of the background superstratum geometry is small, i.e., $N_{00} \sim O(1)$, the tidal force is small in the large $N$ limit
\be
\tilde{A}^{scr}_{m,n}(t) \sim O(N^{-1}) g^2 { t^2 \over m^2} \rightarrow 0, ~~~~~~N_{00}\sim O(1)
\ee
where we have used $g\ll \sqrt{N}$ from (\ref{gravity point}).
When the number is large at order $ O(N)$, the tidal force is not small
\be
\tilde{A}^{scr}_{m,n}(t) \sim O(1) g^2 { t^2 \over m^2},~~~~~~~~N_{00}\sim O(N)
\ee
On the gravity side, when the number of background supergravitions is small, there is only scattering between these background supergravitons and the probe graviton. The scattering is suppressed at large $N$. When the number of background supergravitions is large, they form the superstratum geometry and the probe graviton can be easily tidally excited into stringy states.

\section{Discussion}\label{sec 7}

In this paper we have investigated the tidal excitation of a graviton probe moving within a $(1,0,n)$ superstratum geometry from the perspective of the dual D1D5 CFT. In the gravity computation \cite{mw,chl} the authors found that the kinetic energy of the probe along the radial direction was converted into string excitations. 
In the CFT the graviton we considered is made from a superposition of a left and right moving bosonic excitations.
To study these tidal excitations we have turned on a deformation of the CFT which includes a twist operator. This is the basic interaction of the D1D5 system which joins and splits `component strings' or `strands' created by the bound state of D1 and D5 branes wrapping the common $S^{1}$ direction. 

In particular, we have used two deformation operators to compute the transition amplitude for one left and one right moving mode in the initial state to split into three left and three right moving modes in the final state. We computed this process in the background of a $(1,0,n)$ superstratum state which we labeled as the CFT dual of the $(1,0,n)$ superstratum geometry. 
We found that there is a $t^2$ growth in the amplitude. This is because through the deformation operator, the probe is allowed to interact with the left and right moving modes in the superstratum state. This allows the initial probe to split into more modes with lower energies in the final state, a  process suggestive of a graviton being tidally excited within the supergravity picture. This is in contrast to the CFT dual of a graviton probe moving within empty AdS \cite{hm,dissertation,Guo:2021ybz}. The amplitude of this latter process was found to only oscillate with a period of $2\pi$. This is consistent with the fact that a probe in empty AdS moving along a geodesic will not be tidally excited or thermalized but only redshifted.  

In this paper, we have studied the simplest splitting process in the background of the $(1,0,n)$ superstratum state. We believe that the $t^2$ behavior, which signal tidal forces, is generally true for other types of initial and final states. These correspond to other types of initial probes and final stringy states in other backgrounds. For example, we would like to consider similar splitting processes within a background state in the long winding sector of the CFT which is believed to correspond to a long throat in the gravity dual. Furthermore, the $t^2$ behavior appearing in the splitting amplitude originates from intermediate states which carry the same energy as the initial and final states \cite{hm}. Therefore by restricting to these set of states it may be possible to derive the $t^2$ term analytically.
This could be done by studying the effect of only one deformation operator instead of computing the amplitude with two deformation operators. Here, many of the expressions are known analytically in closed form. 
It would be interesting to investigate the CFT dual of the backreaction of the geometry due to the probe. For example, what do these splitting amplitudes look like if the superstratum momentum is absorbed by the probe or transferred to the probe or what happens at long timescales after the probe has been tidally excited many times and is absorbed by the superstratum geometry.  
We hope to return to these in future works.

\section{Acknowledgements}
We would like to thank Iosif Bena, Nejc Čeplak, Yixuan Li, Samir Mathur and Nick Warner for helpful discussions. The work of B.G. is supported by DOE Grant DE-SC0011726. The work of S.H. is supported by the ERC Grant 787320 - QBH Structure.

\end{document}